\documentclass[%
aip,
 amsmath,amssymb,
 reprint,
]{revtex4-1}

\usepackage{graphicx}
\usepackage{dcolumn}
\usepackage{bm}
\usepackage{xcolor}
\usepackage{braket}
\usepackage{hyperref}
\newcommand\etal{\mbox{\textit{et al. }}}

\draft 

\begin{document}

\title{Spatially heterogeneous dynamics and locally arrested density fluctuations from first-principles.}

\author{J. Lira-Escobedo}
\affiliation{Instituto de F\'{\i}sica ``Manuel Sandoval Vallarta", Universidad Aut\'{o}noma de San Luis Potos\'{i}, \' Alvaro Obreg\' on 64, 78000, San Luis Potos\' i, M\' exico.}

\author{J.R. V\'elez-Cordero}
\affiliation{Investigadores CONACYT-Instituto de F\'{\i}sica ``Manuel Sandoval Vallarta", Universidad Aut\'{o}noma de San Luis Potos\'{i}, \' Alvaro Obreg\' on 64, 78000, San Luis Potos\' i, M\' exico.}

\author{Pedro E. Ram\'{\i}rez-Gonz\'{a}lez}
\email{pramirez@ifisica.uaslp.mx}
\affiliation{Investigadores CONACYT-Instituto de F\'{\i}sica ``Manuel Sandoval Vallarta", Universidad Aut\'{o}noma de San Luis Potos\'{i}, \' Alvaro Obreg\' on 64, 78000, San Luis Potos\' i, M\' exico.}

\date{\today}

\begin{abstract}
We present a first-principles formalism for studying dynamical heterogeneities in glass forming liquids. Based on the Non-Equilibrium Self-Consistent Generalized Langevin Equation theory, we were able to describe the time-dependent local density profile during the particle interchange among small regions of the fluid. The final form of the diffusion equation contains both, the contribution of the chemical potential gradient written in terms of a coarse-grained density and a collective diffusion coefficient as well as the effect of a history-dependent mobility factor. With this diffusion equation we captured interesting phenomena in glass forming liquids such as the cases when a strong density gradient is accompanied with a very low mobility factor attributable to the denser part: in such circumstances the density profile falls into an arrested state even in the presence of a density gradient. On the other hand, we also show that above a certain critical temperature,which depends on the volume fraction, any density heterogeneity relaxes to a uniform state in a finite  time, known as equilibration time. We further show that such equilibration time varies little with the temperature in diluted systems but can change drastically with temperature in concentrated systems.
\end{abstract}

\maketitle 

\section{Introduction}

The study of structural heterogeneities constitute a common area out of many different processes occurring near or far from equilibrium. From the technological viewpoint they are important, for example, in the stability of bubbly flows \cite{Hoefsloot1993,Tyagi2017}, food products \cite{deKruif2012,Zhu2019}, water purification processes involving coagulation and flocculation \cite{Wilts2018} and in the transport of crude oil along the extraction pipes avoiding asphaltene deposition \cite{Alhosani2020}, to mention a few. On the other hand, spatial and dynamical heterogeneities have been observed in well controlled laboratory experiments or numerical simulations involving the formation of dendritic structures \cite{Ratka2019}, spontaneous emulsification \cite{Solans2016} or in glass-forming liquids near its glass transition temperature or $T_g$ \cite{Wang2019}. Among the several attempts that have been made in embracing some of these phenomena in a common foundational theoretical description, the Dynamic Density Functional Theory (DDFT) arises as one of the most useful theoretical frameworks \cite{wittkowski2011dynamical}. Classical Density Functional Theory (DFT) has been widely applied to study equilibrium properties of uniform and homogeneous fluids \cite{wu2007density} and it has helped to understand the complex behavior of non-uniform liquids as well\cite{evans1979nature,evans1992density} . The dynamic extension of DFT, initially based on phenomenological considerations \cite{dieterich1990nonlinear} and later derived from first principles \cite{marconi2000dynamic,marconi1999dynamic,archer2004dynamical}, has allowed to include time-dependent problems for the evolution of the density profile. This dynamic formalism of DFT allows the study of slow dynamics in liquids but it is unable to describe the rapid increase of the relaxation time in supercooled liquids \cite{wu2007density}. On the other hand, taking into account DFT arguments together with some other microscopic considerations, Random First Order Transition (RFOT) theory has emerged as an alternative framework that has already proved to be very useful in the context of glass-transitions\cite{lubchenko2004theory,lubchenko2007theory}.

Another existent theoretical framework for time-dependent problems that allows the study of glass transitions is the Non-Equilibrium Self-Consistent Generalized Langevin Equations (NE-SCGLE) theory \cite{pedro1,pedro2}. Such a formalism is based on functional derivatives (or correlation functions) and its most general version can describe the dynamics and kinetics of non-uniform systems by solving a closed set of time-dependent equations for the density profile, correlations functions and particle mobility. The NE-SCGLE theory, like RFOT, starts with a proper Helmholtz free energy defining the interparticle interaction potential and then proceeds with the computation of the structural evolution of liquids under relaxation processes. It has been shown, for example, that within the NE-SCGLE treatment, dynamic heterogeneity and aging come naturally from solving such a closed set of equations \cite{pedro1,lira2019first,lira2021fundamental}. In contrast, for other formalism such as RFOT, these features of glasses are only observed if one introduces strong correlations between particles \cite{lubchenko2007theory}. In Lira-Escobedo \etal \cite{lira2019first,lira2021fundamental} we have addressed some qualitative difference and similarities between the NE-SCGLE theory and RFOT. Unfortunately, solving the equations belonging to the most general version of the NE-SCGLE theory is a difficult task and so far, it has been solved for  simplified but illustrative problems  \cite{olais,rivas2018different,lira2019first,paty2017,pedro2,LuisEnrique2013,olais2019interference}. In particular, the NE-SCGLE theory has been applied in uniform systems under isochoric conditions and applying temperature jumps (instantaneous change) to study the structural relaxation. The equilibrium counterpart of this simplified version of the NE-SCGLE theory, known as SCGLE theory, shares several similarities with the Mode Coupling Theory (MCT) \cite{elizondo2019glass}. However, the non-equilibrium extension of this theory has demonstrated to go beyond the limits of MCT\cite{lira2021fundamental}. A thoroughly comparison between MCT and SCGLE and an analysis addressing the differences between MCT and NE-SCGLE theories can be found elsewhere \cite{elizondo2019glass,lira2021fundamental}. The simplified version of the NE-SCGLE theory has also been extended to include cooling rate effects with an exponential approach \cite{lira2019first} and also it has been extended to study the relaxation of the fictive temperature and therefore the Kovacs’ kinetics glass transition signatures\cite{lira2021fundamental}. 

The aim of this work is to propose a numerical method to solve the most general version of the NE-SCGLE theory capable to compute the relaxation of the density profile, including the possibility of dynamic arrest, in a system that is non-uniform or heterogeneous from the beginning and which is subject to an instantaneous quench. These initial heterogeneities can appear, in turn, by small wavelength fluctuations of the density profile induced by the thermal bath or by long wavelength disturbances given a certain artificial preparation. In this way, the long-time heterogeneity is seen as a consequence of the slow dynamics of the density correlation functions. Our formalism is based on the functional Taylor's expansion of the chemical potential introduced by Evans \etal \cite{evans1979nature}. Using this procedure, we were able to reproduce well-known results of the non-linear diffusion equation such as the linear dependence of the chemical potential on the density gradient and the dependence of the density time derivative on the collective diffusion coefficient \cite{dieterich1990nonlinear,dieterich1980theoretical,sadighi2007exact}. In order to apply the theoretical framework in a concrete example, we chose to study the diffusion of confined soft-spheres under isothermal conditions. We demonstrate that the density-relaxation has a strong dependence with the initial density-profile and temperature. In this sense, we find that a system, initially containing arrested zones, is unable to relax towards a state with uniform density even in the presence of density gradients. Our method to calculate the density-relaxation, given its simplicity, could be expanded to study more complicated and realistic situations such as including cooling rates or considering fluxes of heat or matter.

The paper is organized as follows: In Sec. \ref{sec:a_simplified_model} we introduce a simplified model to study non-uniform systems. A local chemical state equation, needed in our picture of non-uniform systems, is derived in Sec. \ref{sec:local_chemical_state}. In Sec. \ref{sec:calculation_of_the_local} we derive a method to calculate the local time-dependent mobility for a non-uniform system. In Sec. \ref{sec:a_fundamental_model} we finally apply our formalism to study the relaxation of the particle density under isothermal conditions. Lastly, the conclusions and the numerical methods of this work are presented in Sec. \ref{sec:conclusions} and in App. \ref{appendix}.

\section{A simplified picture of a non-homogeneous fluid and the NE-SCGLE equations.}\label{sec:a_simplified_model}

\begin{figure}
\includegraphics[width=0.45\textwidth]{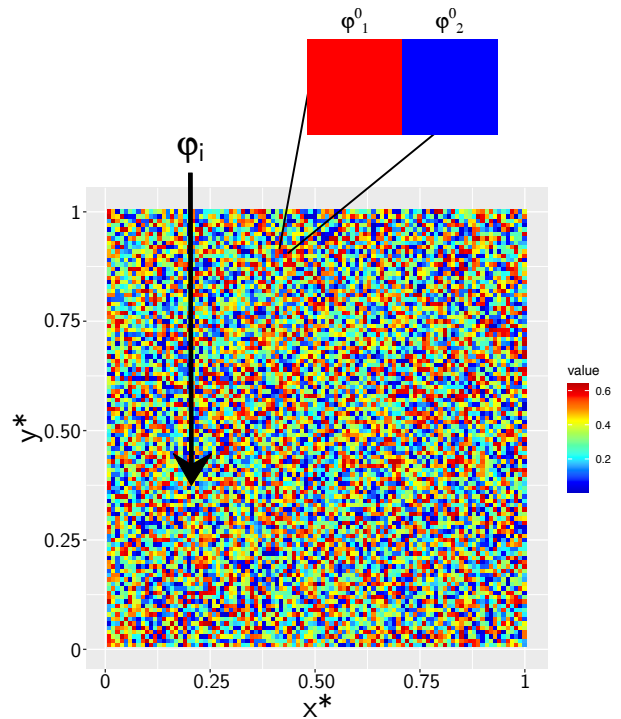}
\caption{Color map of the volume fraction (particle density) of a heterogeneous system. Each cell has a volume fraction of $\varphi _i$. The amplification shows two adjacent cells each of them with a volume fraction of $\varphi ^0 _1$ and $\varphi ^0 _2$. $x^*$ and $y ^*$ are defined by $x/L$ and $y/L$, where $L$ is the length of the system. This figure was built using R software \cite{rsoftware}.} 
\label{fig:density_colormap}
\end{figure}

A rather convenient form of visualizing a non-homogeneous fluid is to consider a two dimensional patchwork model of a supercooled liquid as shown in Fig. \ref{fig:density_colormap}. Each cell is described by the local density $n_i (t)$ (the figure actually shows the volume fraction $\varphi_i$  instead of the particle density), where $i= 1,2,..., N$, being $N$ the total number of cells and $t$ refers to  the waiting (laboratory) time. Let us assume that the $N$ is large enough so that we can consider the local density as a continuum variable. Thus, the local density $n _i(t)$ becomes $n(\mathbf{r},t)$, where $\mathbf{r}$ is the position vector. In what follows we present the NE-SCGLE formalism in order to describe such a model for heterogeneous fluids.

The most general version of the NE-SCGLE theory \cite{pedro1} allows the description of the relaxation of the mean value of the density profile $\bar{n}(\mathbf{r},t)$ as well as its fluctuations around this mean using the non-uniform single time matrix covariance $\sigma(k;\mathbf{r},t)$ and the  non-uniform correlation functions $C(k,\tau;\mathbf{r},t)$ and $C _S(k,\tau;\mathbf{r},t)$, where $\bar{n}(\mathbf{r},t)$, $\sigma(k;\mathbf{r},t)$, $C(k,\tau;\mathbf{r},t)$ and $C _S(k,\tau;\mathbf{r},t)$ are coupled by the non-uniform time-dependent mobility $b(\mathbf{r},t)$ and the non-uniform time-dependent irreducible memory function $\Delta \zeta (\tau;\mathbf{r},t)$, where $k$ refers to the Fourier space. NE-SCGLE theory further assumes that the system is isotropic, thus, the two-point correlation functions depend only on the distance between such points, $|\mathbf{x} - \mathbf{x} ^{\prime}|$. In this way, the inverse Fourier transform of the two-point  correlation functions become $\sigma(|\mathbf{x} - \mathbf{x} ^{\prime}|;\mathbf{r},t) = \overline{\delta n(\mathbf{r} + \mathbf{x}, t)\delta n(\mathbf{r} + \mathbf{x} ^{\prime},t)} $, $C(|\mathbf{x} - \mathbf{x} ^{\prime}|,\tau;\mathbf{r},t) = \overline{\delta n(\mathbf{r} + \mathbf{x}, t + \tau)\delta n(\mathbf{r} + \mathbf{x} ^{\prime},t)} $ and its self component $C _S(|\mathbf{x} - \mathbf{x} ^{\prime}|,\tau;\mathbf{r},t)$ calculated as

\begin{equation}
f (| \mathbf{x} - \mathbf{x ^{\prime}}|) = \frac{1}{(2 \pi)^3} \int d^3ke^{ - i\mathbf{k}\cdot | \mathbf{x} - \mathbf{x ^{\prime}}|}f (k)\;,
\end{equation}

\noindent written for any function $f(x)$. The time-evolution equation for the average density profile, $\bar{n}(\mathbf{r},t)$, is given by the temporally and spatially local but non-linear diffusion equation \cite{pedro1}

\begin{equation}
\frac{\partial \bar{n}(\mathbf{r},t)}{\partial t} = D _0 \nabla \cdot \bigg\{ \bar{n}(\mathbf{r},t)b(\mathbf{r},t)\nabla \bigg[\beta \mu[\mathbf{r}; \bar{n}(\mathbf{r},t)] \bigg] \bigg\},
\label{eq:diffusion_equation_local}
\end{equation}

\noindent where $D _0$ is the short-time diffusion coefficient and $\beta \mu[\mathbf{r}; \bar{n}(\mathbf{r},t)]$ the chemical potential with the brakets indicating functional dependence on $\bar{n}(\mathbf{r},t)$. The time-equation for the non-uniform single time matrix covariance $\sigma(k;\mathbf{r},t)$ reads as follows:

\begin{align}
\frac{\partial \sigma (k;\mathbf{r},t)}{\partial t} & = -2 k^2 D_0 \bar{n}(\mathbf{r},t)b(\mathbf{r},t)\mathcal{E}(k;\bar{n}(\mathbf{r},t))\sigma(k;\mathbf{r},t) \nonumber \\
 &+ 2k^2D_0 \bar{n}(\mathbf{r},t)b(\mathbf{r},t),
\label{eq:time-evolution-covariance}
\end{align}

\noindent where $\mathcal{E}(k;\bar{n}(\mathbf{r},t)) \equiv \mathcal{E}(k;\bar{n}(\mathbf{r},t),T(t))$ (assuming that the temperature profile $T(\mathbf{r}, t)$ remains uniform) is the Fourier Transform of the second functional derivative matrix $\mathcal{E}(|\mathbf{x} - \mathbf{x} ^{\prime}|;\bar{n}(\mathbf{r},t), T (t))$ of the Helmholtz free energy density-functional $\mathcal{F}[n;T]$,

\begin{equation}
\mathcal{E}(|\mathbf{x} - \mathbf{x} ^{\prime}|;\bar{n}(\mathbf{r},t), T (t)) \equiv \bigg( \frac{\delta ^2 \mathcal{F}[n;T/k _B T] }{\delta n (\mathbf{\mathbf{r} + \mathbf{x}})\delta n (\mathbf{\mathbf{r} + \mathbf{x} ^{\prime}})} \bigg) \bigg| _{T,n},
\end{equation}

\noindent evaluated at $n =\bar{n}(\mathbf{r},t)$ and $T = T(t)$. The relaxation of the non-uniform correlation functions $C(k,\tau;\mathbf{r},t)$ and $C _S(k,\tau;\mathbf{r},t)$ are given by:

\begin{align}
\frac{\partial C(k,\tau;\mathbf{r}, t) }{\partial \tau}  &= \lambda(k; \mathbf{r}, t) \Delta \zeta (\tau; \mathbf{r}, t) \sigma (k, \mathbf{r}, t) \nonumber \\
& - \frac{k^2D _0 C(k,\tau;\mathbf{r}, t) \bar{n}(\mathbf{r},t) }{\sigma (k; \mathbf{r}, t)}  \nonumber \\
 -\lambda(k; \mathbf{r}, t) \frac{\partial}{\partial \tau} &\int _{0} ^{\tau} d \tau \Delta \zeta (\tau-\tau^{\prime}; \mathbf{r}, t) C(k,\tau^{\prime}; \mathbf{r}, t)
\label{TimeEvolutionFk},
\end{align}

\begin{align}
\frac{\partial C _S(k,\tau;\mathbf{r}, t) }{\partial \tau}  &= \lambda(k; \mathbf{r}, t) \Delta \zeta (\tau;\mathbf{r}, t) \nonumber \\ 
& - k^2D_0 C _S(k,\tau;\mathbf{r}, t) \nonumber \\
 -\lambda(k; \mathbf{r}, t) \frac{\partial}{\partial \tau} &\int _{0} ^{\tau} d \tau \Delta \zeta (\tau-\tau^{\prime};\mathbf{r}, t) C _S(k,\tau^{\prime}; \mathbf{r}, t)
\label{TimeEvolutionFs},
\end{align}

\noindent in which the function $\lambda(k;\mathbf{r},t)$ is an interpolating function defined by

\begin{equation}
\lambda(k;\mathbf{r},t) = \frac{1}{1 + \bigg( \frac{k}{k _c (\mathbf{r},t) }\bigg) ^2},
\end{equation}

\noindent 
where, $k _c(\mathbf{r}, t) =  ck _{max}(\mathbf{r}, t)$, with $c \in \mathbb{R} $ and $k _{max} (\mathbf{r},t )$ being the main peak of the local time-dependent structure factor $S(k;\mathbf{r},t) = \sigma(k;\mathbf{r},t) /\bar{n}(\mathbf{r};t)$ at $\mathbf{r}$ and $t$. Parameter $c$ has been set to 1.109 in all the studies using NE-SCGLE in order to reproduce the arrest scenario of hard spheres \cite{juarez2007simplified}.

The irreducible non-uniform memory function $\Delta \zeta(k,\tau;\mathbf{r},t)$ is then computed by: \cite{MedinaFaraday}

\begin{align}
& \Delta \zeta(\tau;\mathbf{r},t) = \frac{D ^0}{3(2 \pi )^3 \bar{n}(\mathbf{r},t)}  \nonumber \\ & \times \int d\mathbf{k} k ^2\bigg[ \frac{\sigma (k;\mathbf{r},t)/\bar{n}(\mathbf{r},t) - 1}{\sigma(k;\mathbf{r},t)}\bigg] ^2 C(k;\tau,\mathbf{r},t)  C _S(k;\tau,\mathbf{r},t).
\label{eq:general_delta_zeta}
\end{align}

Finally, the time-evolution of the non-uniform mobility $b (\mathbf{r},t)$ is given by:

\begin{equation}
b(\mathbf{r},t) = \frac{1}{1 + \int _0 ^{\infty} d \tau \Delta \zeta (\tau;\mathbf{r},t)}.
\label{eq:local_mobility}
\end{equation}

Eqs. (\ref{eq:diffusion_equation_local}) - (\ref{eq:local_mobility}) constitute a closed set of equations for the dynamics and kinetics of a non-uniform system and they are the most general version of the NE-SCGLE theory for colloidal dynamics. The only external input is the chemical state equation $\beta \mu[\mathbf{r}; \bar{n}(\mathbf{r},t)] $. A complete description of the theory discussed here can be found elsewhere \cite{pedro1}. 

Notice that Eq. (\ref{eq:diffusion_equation_local}) reduces to the mean-evolution equation of the dynamic DFT when the mobility $b(\mathbf{r},t)$ is taken as constant \cite{dieterich1990nonlinear,marconi2000dynamic,marconi1999dynamic,archer2004dynamical}. Let us emphasize that in both, the dynamic DFT and in the NE-SCGLE theory, the local and non-linear equation for the density profile is obtained starting from the Langevin equation \cite{pedro1}. Thus, in principle, the NE-SCGLE theory extends the picture of DFT for colloidal dynamics by including a non-uniform time-dependent mobility.

\section{Local chemical state equation and local diffusion equation}\label{sec:local_chemical_state}
\noindent
The most general state chemical equation of a heterogeneous fluid can be written as:\cite{pedro1}

\begin{equation}
\beta \mu[\mathbf{r}, n(\mathbf{r},t)] = \beta \mu ^* (\beta) + \mbox{ln} n(\mathbf{r},t) - c[\mathbf{r}; n(\mathbf{r},t)] + \beta \psi(\mathbf{r}),
\label{most_general_chemical_equation}
\end{equation}

\noindent where $\psi(\mathbf{r}) $ is the external potential acting on the particles at position $\mathbf{r}$. The terms $ \beta \mu ^* (\beta) + \mbox{ln} n(\mathbf{r},t) $ are the ideal-gas contribution to the chemical potential and $- c[\mathbf{r}; n(\mathbf{r},t)]$ represents the deviations of the ideal gas due to the interparticle interactions. Thus, in the absence of external fields, we can expand $\beta \mu [\mathbf{r}, n(\mathbf{r},t)] $ as a functional Taylor's series  around a uniform density $n_u$, namely

\begin{multline}
\beta \mu[\mathbf{r}, n(\mathbf{r},t)] = \beta \mu [n _u] + \int d \mathbf{r ^{\prime}} \frac{\delta \beta \mu [n(\mathbf{r},t)]}{\delta n(\mathbf{r ^{\prime}},t)} \bigg| _{n _u} \tilde{n}(\mathbf{r ^{\prime}},t) \\ + \cdots,
\label{taylor_series}
\end{multline}

\noindent where $\tilde{n}(\mathbf{r},t) = n( \mathbf{r},t ) - n_u$ is the deviation  from the uniform density. Keeping up to linear terms in the expansion and deriving Eq. (\ref{most_general_chemical_equation}) with respect to  $n(\mathbf{r},t)$ to compute the derivative inside the integral of Eq. (\ref{taylor_series}), we obtain that:

\begin{multline}
\beta\mu [\mathbf{r}; n(\mathbf{r},t)]] = \beta \mu [n _u]  +\\ \int d \mathbf{r} ^{\prime} \bigg[ \frac{\delta(|\mathbf{r} - \mathbf{r ^{\prime}}|)}{n _u} - c[\mathbf{r}, \mathbf{r ^{\prime}}; n_u ]\bigg]  \tilde{n}(\mathbf{r ^{\prime}},t),
\label{chemical_potential_r}
\end{multline}

\noindent
where $c[\mathbf{r}, \mathbf{r^{\prime}}; n _u]$ is the Ornstein–Zernike direct correlation function \cite{mcquarrie}. Defining $\mathbf{r} = \mathbf{r} + \mathbf{x}$ and $\mathbf{r} ^{\prime} = \mathbf{r} + \mathbf{x} ^{\prime}$ and assuming that the system is isotropic we have that:

\begin{multline}
\beta\mu [\mathbf{r}, \mathbf{x}; n(\mathbf{r},t)] = \beta \mu [n _u] +  \\ \int d \mathbf{x} ^{\prime} \bigg[ \frac{\delta(|\mathbf{x} - \mathbf{x ^{\prime}}|)}{n _u} - c[|\mathbf{x}-\mathbf{x ^{\prime}|; \mathbf{r}, n_u} ]\bigg] \tilde{n}(\mathbf{x ^{\prime}},\mathbf{r},t),
\label{chemical_potential_x}
\end{multline}

\noindent
where $\tilde{n}(\mathbf{x ^{\prime}},\mathbf{r},t) = n(\mathbf{r} + \mathbf{x} ^{\prime},t) - n _u$. Thus, in order to calculate the chemical potential at $\mathbf{r}$-position we need to integrate over the $\mathbf{x}$ coordinate: 

\begin{equation}
\beta\mu [ \mathbf{r}; n(\mathbf{r},t)] = \frac{1}{V}\int _ {V }d\mathbf{x} \beta\mu [\mathbf{r}, \mathbf{x}; n(\mathbf{r},t)],
\label{V_integral}
\end{equation}

being $V$ the volume of the system.

Introducing Eq. (\ref{chemical_potential_x}) into Eq. (\ref{V_integral}) we therefore obtain that:

\begin{multline}
\beta\mu [n(\mathbf{r},t)] = \frac{\beta \mu [n _u] }{V}\int _ {V}d\mathbf{x} + \\ \frac{\tilde{n}(\mathbf{r},t)}{V} \int _ {V}\int _ {V}d\mathbf{x}  d \mathbf{x ^{\prime}} \bigg[ \frac{\delta(|\mathbf{x} - \mathbf{x ^{\prime}}|)}{n _u} - c[|\mathbf{x}-\mathbf{x} ^{\prime}|; n_u ]\bigg], 
\label{chp}
\end{multline}

\noindent where we have assumed that $\tilde{n}(\mathbf{x ^{\prime}},\mathbf{r},t) \approx \tilde{n}(\mathbf{r},t)$ and neglected the dependence of $c[|\mathbf{x}-\mathbf{x} ^{\prime}|; n_u]$ on $\mathbf{r}$ since it is evaluated at $n _u$. It is important to remark that the first assumption considers that the fluctuation differences in the vicinity of $\mathbf{r}$ are so small that they can be approximated up to the zero order term or just considering the position centered at $\mathbf{r}$. Now, the second term of the last equation can be written as

\begin{multline}
\int _ {V}\int _ {V}d\mathbf{x}  d \mathbf{x} ^{\prime} \bigg[ \frac{\delta(|\mathbf{x} - \mathbf{x ^{\prime}}|)}{n _u} - c[|\mathbf{x}-\mathbf{x} ^{\prime}|; n_u ]\bigg] = \\  
V \int _ {V}d\mathbf{y}  \bigg[ \frac{\delta(|\mathbf{y}|)}{n _u} - c[|\mathbf{y}|; n_u]\bigg],
\label{eq:integral}
\end{multline}

\noindent making the substitution $\mathbf{y} = \mathbf{x} - \mathbf{x ^{\prime}}$. The integral in $\mathbf{y}$ can be readily found using the properties of the Fourier transform:

\begin{equation}
\int _ {V }d\mathbf{y}  \bigg[ \frac{\delta(|\mathbf{y}|)}{n _u} - c[|\mathbf{y}|; n_u]\bigg] = \frac{1}{n _u} - C ^{eq}(k=0; n_u),
\end{equation}

\noindent
where, for any value of k, the general expression reads as:

\begin{equation}
\frac{1}{n _u} - C ^{eq}(k; n_u) = \int _ {V}d\mathbf{y} e^{-i\mathbf{k}\cdot |\mathbf{y}|} \bigg[ \frac{\delta(\mathbf{y})}{n _u} - c[n _u,y]\bigg].
\end{equation}

\noindent
The function $C ^{eq}(k; n_u)$ is related to the equilibrium structure factor \cite{mcquarrie} by $ S^{eq}(k; n_u) = (1 - n _u C ^{eq}(k; n_u))^{-1} $, thus

\begin{equation}
\frac{1}{n _u} - C ^{eq}(k=0; n_u) = \frac{1}{n _u S^{eq}(k=0; n_u)}.
\label{eq.structure}
\end{equation}

\noindent
Combining Eqs. (\ref{chp}) to (\ref{eq.structure}) we have that 

\begin{equation}
\beta\mu [n(\mathbf{r},t)] = \beta \mu [n _u]  + \frac{1}{n _u S^{eq}(k=0; n_u)}\tilde{n}(\mathbf{r},t).
\label{chp_local}
\end{equation}

\noindent
With the result of the Eq. (\ref{chp_local}) we have that the diffusion equation reads as follows:

\begin{equation}
\frac{\partial \bar{n}(\mathbf{r}, t)}{\partial t} = D ^* _0  \nabla \cdot [ b(\mathbf{r},t) \bar{n}(\mathbf{r}, t) \nabla  \bar{n}(\mathbf{r}, t) ], 
\label{Difussion_equation}
\end{equation}

\noindent
where $D _0 ^* = D _0 / n _u S^{eq}(k=0; n_u) = D _ {coll} (\mathbf{r})/n _u$, being $D _ {coll} (\mathbf{r}) = D _0 / S^{eq}(k=0; n_u) $ the collective diffusion coefficient that describe long-wavelength density fluctuations behavior \cite{dieterich1980theoretical}.

It is instructive to compare the present diffusion expression, Eq. (\ref{Difussion_equation}), with other theoretical views that goes from the simple Fickian expression and passing through the Onsager phenomenological coefficients up to the well-known Maxwell-Stefan diffusion for multicomponent systems \cite{Leonardi2010}. As discussed by Wang and LeVan \cite{Wang2008}, all these three theoretical descriptions are somehow equivalent and their selection is more a matter of applicability (evidently, Fick's diffusion coefficient is, for example, reducible in the sense that it can be further decomposed in terms including Maxwell-Stefan or Onsager coefficients). Comparing with the present formulation, it is evident that it also shares elements with these other expressions and equivalences can be defined among them. Notice that Eq. (\ref{Difussion_equation}) has exactly the same form of the the standard non-linear diffusion equation for the relaxation of density with diffusion coefficient $D (\bar{n} (\mathbf{r},t)) = D ^* _0(\mathbf{r}) b(\mathbf{r},t)\bar{n} (\mathbf{r},t)$\cite{dieterich1990nonlinear,sadighi2007exact}. Also, the present formulation uses the gradient of the chemical potential, as it is done in the Onsager and Maxwell-Stefan expressions, and also contains a non-local mobility which is analogous to the friction terms capture in Maxwell and Stefan's analysis. In any case, although the present approach was done for monocomponent systems, it has the advantage that is free of phenomenological coefficients and its components came directly from the microestructure and interparticle interactions information.

\noindent
Eq. (\ref{Difussion_equation}) together with the set of Eqs. (\ref{eq:time-evolution-covariance})-(\ref{eq:local_mobility}) represent a simplified version of the NE-SCGLE theory for the relaxation of the average of the density profile , $\bar{n}(\mathbf{r}, t)$, in a heterogeneous systems. However, the solution of those equations is a difficult task and we require some other approximations.

\section{Calculation of the local time-dependent mobility.}\label{sec:calculation_of_the_local}

Let us mention that $\bar{n}(\mathbf{r}, t)$ along with its fluctuations $\sigma (k, \mathbf{r}, t)$ describe the overall macroscopic relaxation in the sense that if we know $b(\mathbf{r},t)$ for all $\mathbf{r}$ and $t$, we do not need to solve for the microscopic information provided by Eqs. (\ref{TimeEvolutionFk}) - (\ref{eq:local_mobility}). As a first step, lets assume that we are only interested in the relaxation of $\bar{n}(\mathbf{r}, t)$. Furthermore, lets assume that any historical relaxation path of the fluctuations $\sigma (k, \mathbf{r}, t)$ visit, in their sequence of transitions, the exactly same values as its equilibrium counterparts obtained in an ordered set of equal $\bar{n}$'s (\textit{i.e.}, we assumed instantaneous local relaxation for the fluctuations). Formally, we can decouple  $\sigma (k, \mathbf{r}, t)$ from the others NE-SCGLE equations by writing $\sigma (k, \mathbf{r}, t) = \sigma ^{eq} (k; \bar{n}(\mathbf{r}, t)) + \Delta \sigma (k, \mathbf{r}, t) $, where $\Delta \sigma (k, \mathbf{r}, t) \equiv \sigma (k, \mathbf{r}, t)  - \sigma ^{eq} (k; \bar{n}(\mathbf{r}, t))$, and then neglect the non-equilibrium effects provided by $\Delta \sigma (k, \mathbf{r}, t)$. With this assumption we no longer need the time equation for $\sigma (k, \mathbf{r}, t)$. In particular, we can write Eq. (\ref{Difussion_equation}) as:

\begin{equation}
\frac{\partial \varphi(\mathbf{r}, t)}{\partial t} =  D ^* _0\nabla \cdot [ b(\varphi(\mathbf{r},t)) \varphi(\mathbf{r}, t) \nabla \varphi(\mathbf{r}, t) ],
\label{Difussion_equation_eq}
\end{equation}

\noindent where $\varphi(\mathbf{r},t)$ is  the time-dependent volume fraction, $\varphi(\mathbf{r},t) = \pi a ^3 n(\mathbf{r},t)/6$ for spheres of radius $a$. Henceforth, we focus on the relaxation of the local time-dependent mobility $b(\varphi(\mathbf{r},t))$ which can be calculated by solving the equations for the equilibrium intermediate scattering functions, $F ^{eq}(k,\tau;[\varphi(\mathbf{r},t)]), F ^{eq}_{S}(k,\tau;[\varphi(\mathbf{r},t)])$ and the equilibrium memory kernel $\Delta \zeta ^{eq} (\tau;[\varphi(\mathbf{r},t)])$ given by:

\begin{multline}
\frac{\partial F ^{eq}(k,\tau;[\varphi(\mathbf{r},t)])}{\partial \tau}  = \lambda(k) \Delta \zeta ^{eq} (\tau; [\varphi(\mathbf{r},t)]) S ^{eq}(k,[\varphi(\mathbf{r},t)]) \\ - \frac{k^2D_0 F ^{eq}(k,\tau; [\varphi(\mathbf{r},t)])}{S ^{eq}(k,[\varphi(\mathbf{r},t)])} \\ -\lambda(k)\frac{\partial}{\partial \tau} \int _{0} ^{\tau} d \tau \Delta \zeta ^{eq} (\tau-\tau^{\prime} ;[\varphi(\mathbf{r},t)]) F ^{eq}(k,\tau^{\prime};[\varphi(\mathbf{r},t)]),
\label{TimeEvolutionFk_eq}
\end{multline}

\begin{multline}
\frac{\partial F ^{eq}_{S}(k,\tau;[\varphi(\mathbf{r},t)])}{\partial \tau}  = \lambda(k) \Delta \zeta ^{eq} (\tau; [\varphi(\mathbf{r},t)])  \\ - k^2 D_0 F ^{eq} _{S}(k,\tau; [\varphi(\mathbf{r},t)]) \\ -\lambda(k) \frac{\partial}{\partial \tau} \int _{0} ^{\tau} d \tau \Delta \zeta ^{eq} ( \tau-\tau^{\prime}; [\varphi(\mathbf{r},t)]) F ^{eq}_{S}(k,\tau^{\prime}; [\varphi(\mathbf{r},t)]),
\label{TimeEvolutionFs_eq}
\end{multline}

\begin{multline}
\Delta \zeta ^{eq} (\tau;[\varphi(\mathbf{r},t)]) = \frac{D_0}{6 \pi ^2 n(\mathbf{r},t)} \int _{0}^{\infty} dk k^4 \times \\ \bigg[ \frac{S ^{eq}(k;[\varphi(\mathbf{r},t)]) -1}{S ^{eq}(k;[\varphi(\mathbf{r},t)])}\bigg]^2\\
\times F ^{eq}(k,\tau;[\varphi(\mathbf{r},t)])F ^{eq} _{S}(k,\tau;[\varphi(\mathbf{r},t)]),
\label{TimeEvolutiondzeta_eq}
\end{multline}

\noindent where $ S ^{eq}(k,[\varphi(\mathbf{r},t)]) $ is the equilibrium structure factor and the brakets $[\dots]$ inside $F ^{eq}$, $F ^{eq} _S$, $\Delta \zeta ^{eq}$ and $S ^{eq}$ mean functional dependence on the local volume fraction $\varphi(\mathbf{r},t)$. $\lambda(k;\mathbf{r},t)$ is again given by:

\begin{equation}
\lambda(k;\mathbf{r},t) = \frac{1}{1 + \bigg( \frac{k}{k _c (\mathbf{r},t) }\bigg) ^2},
\label{eq:lambdak}
\end{equation}

\noindent where now $k _c (\mathbf{r},t) = c k _{max} (\mathbf{r},t)$ is defined in terms of the $S ^{eq}(k;\varphi(\mathbf{r},t))$ profile (here again $c = 1.109$ to reproduce the dynamics of hard-spheres \cite{juarez2007simplified}). The mobility is obtained by:

\begin{equation}
b(\varphi(\mathbf{r},t)) = \bigg[ 1 + \int _{0} ^{\infty} d\tau \Delta \zeta (\tau;[\varphi(\mathbf{r},t)]) \bigg]^{-1}
\label{TimeEvolutionmovility_eq}.
\end{equation}

\noindent
Let us emphasize that the zero order approximation of $\sigma (k, \mathbf{r},t)$ does not mean that the fluctuations are not important in the relaxation of the density $\bar{n}(\mathbf{r},t)$. In fact, the fluctuations  $\sigma ^{eq} (k, \bar{n} (\mathbf{r},t)) = S ^{eq}(k;\bar{n}(\mathbf{r},t)) / \bar{n}(\mathbf{r},t) $ enter in our picture through the equilibrium mobility $b(\bar{n}(\mathbf{r},t))$ which contains the information of the interparticle forces. 
Thus, our approach introduces a self-consistent method to describe the relaxation of the density and the equilibrium mobility. The self-consistent scheme provided by NE-SCGLE theory is the main difference with other approaches, for instance within dynamic DFT \cite{marconi1999dynamic,marconi2000dynamic} where a constant mobility is assumed. On the other hand, analytic functions for the local diffusion equations are required to apply nonlinear diffusion equations \cite{dieterich1990nonlinear, sadighi2007exact}.

\noindent
Given the structure of the Eqs. (\ref{Difussion_equation_eq}) - (\ref{TimeEvolutionmovility_eq}) we can calculate separately the equilibrium mobility $b(\varphi)$ for any value of the density $\varphi$ and then use interpolation formulas to compute $b(\varphi)$ inside Eq.  (\ref{Difussion_equation_eq}) and solve for the time evolution of $\varphi (\mathbf{r},t)$. Given a initial condition, $\varphi(\mathbf{r},t=0)$, we can built an initial mobility profile  $b(\varphi(\mathbf{r},t=0))$ and then use Eq. (\ref{Difussion_equation_eq}) to calculate the change in the density for the marching time $t>0$. In App. \ref{appendix} we provide a numerical method to solve Eq. (\ref{Difussion_equation_eq}).

 \section{A fundamental model of dynamical heterogeneties: isothermal density relaxation of soft spheres.}\label{sec:a_fundamental_model}

In order to develop a fundamental picture of the dynamical heterogeneties in glass-forming liquids, let us consider again the two dimensional patchwork model of a supercooled liquid as shown in Fig. \ref{fig:density_colormap}. At this point we have all the theoretical elements required for the description of the relaxation process of the local volume fraction $\varphi (\mathbf{r},t)$. In order to obtain quantitative results, let us consider two  adjacent  cells in the patch work model of Fig \ref{fig:density_colormap}, each of them characterized by the initial local volume fraction $\varphi ^0 _1$ and $\varphi ^0 _2$ respectively, like the ones showed in the amplification of Fig. \ref{fig:density_colormap}. Hence, rather than describe all the diffusion throughout the patchwork, we describe in detail the diffusion along two adjacent cells; this process can be further spread over all the system. In this manner we can analyze in a simpler way the main predictions of the NE-SCGLE theory.
 
Let us apply the numerical scheme shown in App. \ref{appendix} (Eq. (\ref{discrete_form})) to a one-dimensional heterogeneous  mono-component  fluid made of soft spheres of diameter $a$ immersed in a solvent and interacting by the 6-12 ($\nu=6$) truncated Lennard-Jones potential, (6-12TLJP) \textit{i.e.}, $ \beta u(r) = \epsilon [(\frac{a}{r})^{2\nu} - 2 ( \frac{a}{r} )^{\nu} + 1]  $ for $ r\leq a$ and $\beta u(r) = 0 $ for $r\geq a$. In order to calculate the structure factor for this potential we used the equivalence between soft spheres and hard spheres, $S^{eq}_{SS}(k a;\varphi,T,\nu) = S ^{eq}_{HS}(\lambda_{1} k a;\lambda_{1} ^3 \varphi)$, where $\lambda_{1}$ is computed with the blip function\cite{hansen1990theory}, $\lambda_{1}( T ^*, \nu) = \bigg\{ 1 - 3 \int _0 ^1 dx x^2 \mbox{exp}\bigg[ -\frac{1}{ T^{*}}(\frac{1}{x ^{2 \nu}} - \frac{2}{x ^{\nu}} + 1)\bigg]  \bigg\}$  with $T ^*$ being the reduced temperature.  The structure factor  $S ^{eq}_{HS}(\lambda_{1} k a;\lambda_{1} ^3 \varphi)$ was obtained using the Percus-Yevick closure for the Ornstein-Zernike equation along with the Verlet-Weis approximation for dense liquids\cite{percus1958analysis,verlet1972equilibrium}. The fluid zone and the arrest zone are divided by a critical temperature $T _c^* = T _c ^*(\varphi) $. This critical temperature separates the states that fully equilibrate in any temperature history of those which are unable to equilibrate regardless the thermal history \cite{lira2019first,lira2021fundamental}. The transition temperature $T ^* _c (\varphi)$ was obtained with the equilibrium dynamic arrest criterion derived from an asymptotic analysis of the memory function $\Delta\zeta(\tau, t) $ when $\tau \rightarrow \infty $ \citep{rigo2008prl,sanchezdiaz2009,Rigo2008}. The dynamic arrest criterion reads as follows:

 \begin{figure}
\includegraphics[width=0.4\textwidth]{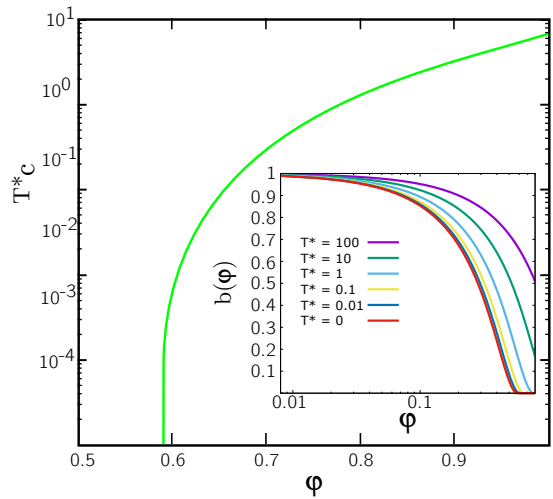}
\caption{Critical temperature $T _c ^*$ as function of the volume fraction $\varphi$. The inset shows the equilibrium mobility $b(\varphi)$ for different temperatures $T ^*$. The point of arrest occurs at $\varphi _c$ where $b(\varphi _c) = 0$.} 
\label{fig:Tc_phi}
\end{figure}

\begin{multline}
\frac{1}{\gamma} = \frac{1}{6 \pi^2 n} \int _0 ^{\infty} dk k^4\\
\times \frac{[S ^{eq}(k;n,T) -1]^2 \lambda ^2 (k)}{[ \lambda (k) S ^{eq}(k;n,T)  + k^2 \gamma][\lambda(k)] + k^2 \gamma]}
\label{gamma},
\end{multline}

\noindent
where $\gamma$ is a localization length. If $\gamma$ is infinite, the system remains in a fluid-like state, whereas if it becomes finite, the system is considered arrested. Fig. \ref{fig:Tc_phi} shows the critical temperature $T _c ^*$ as function of the volume fraction $\varphi$. Inversely, we can also define the critical volume fraction $\varphi _c = \varphi _c (T _c ^*)$ as function of temperature $T ^*$, having $\varphi _c$ the same meaning of $T _c ^*$. The hard-sphere system is obtained in the limit $T^* \rightarrow 0$, giving a value of $\varphi _c= 0.583$, in complete agreement with previous results \cite{perez2011equilibration,brambilla2009probing}.

\noindent
The first input for solving Eq. (\ref{Difussion_equation_eq}) is the equilibrium mobility $b(\varphi)$ which is calculated separately by means of Eqs. (\ref{TimeEvolutionFk_eq}) -  (\ref{TimeEvolutionmovility_eq}). See App. \ref{appendix} for more details in the numerical solution of Eq. (\ref{Difussion_equation_eq}). The inset of Fig. \ref{fig:Tc_phi} shows the equilibrium mobility $b(\varphi)$ as function of the volume fraction $\varphi$ for different temperatures of the system of soft-spheres described above. For the dilute case, given that the interparticle forces are too small, the mobility $b(\varphi)$ is almost temperature independent and always close to unity, \textit{i.e.}, the dynamics are controlled by the short-time diffusion coefficient which only senses the viscosity of the solvent. As the density increases, the mobility $b(\varphi)$ decreases until it reach $b(\phi) = 0$. This condition occurs at $\varphi _c$ and is equivalent to saying that the system is arrested. From the inset of Fig. \ref{fig:Tc_phi} it can be seen that as the temperature increases, the critical volume fraction $\varphi _c$ also increases, as predicted by Eq. (\ref{gamma}) in the main plot of Fig. \ref{fig:Tc_phi}.

Once we have calculated the equilibrium mobility $b(\varphi)$ by solving the discrete form of Eqs. (\ref{TimeEvolutionFk_eq}) -  (\ref{TimeEvolutionmovility_eq}) \cite{Rigo2008}, the numerical solution of Eq. (\ref{discrete_form}) as described in App. \ref{appendix} is straightforward. Let us consider an initial density profile, $\varphi(x ^*,t^* = 0)$, given by the following piece-wise function

\begin{align}
\varphi(x^*, t^* = 0) = \left\{ \begin{array}{cc} 
                \varphi _1 ^0 & \hspace{5mm} 0 \leq x ^* \leq 1/2 \\
                & \\
                \varphi _2 ^0 & \hspace{5mm} 1/2 < x ^* \leq 1\\
                \end{array} \right. , 
\label{initial_condition}
\end{align}

\noindent where $\varphi _1 ^0 < \varphi _2 ^0 < 1$, $t^* = t/\tau _M$, $x^* = x/L$, $L$ is the length of the system and $\tau _M$ is the macroscopic characteristic time defined by $\tau _M = L ^2 /D ^* _0$. Physically, $\tau _M$ is the time required by a tagged particle to diffuse the available volume. We used this configuration as a model for two adjacent cell each with a density $\varphi ^0 _1$ and $\varphi ^0 _2$ (see the amplification in Fig. \ref{fig:density_colormap}). For reference, we refer to the density in $0 < x ^* < 1/2$ as $\varphi _1 (x ^*, t ^*)$ and for $1/2 < x ^* < 1$ as $\varphi _2 (x ^*, t ^*)$. By studying the relaxation of a system with an initial density profile given by Eq. (\ref{initial_condition}), we can understand the relaxation of a heterogeneous system like the one showed in Fig. \ref{fig:density_colormap} since all cell pairs composing the system showed in Fig. \ref{fig:density_colormap} should relax in a similar manner.
 \begin{figure}
\includegraphics[width=0.4\textwidth]{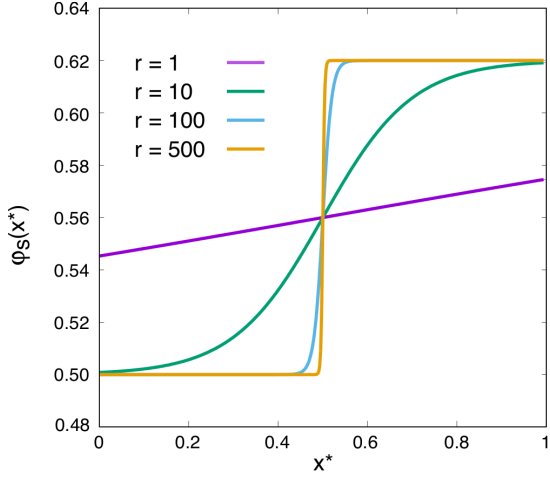}
\caption{ The sigmoid function, $\varphi _s(x^*)$, is plotted for several values of $r$. As $r$ increases, the sigmoid function $\varphi _s(x^*)$ approaches to $\varphi (x^*, t^*=0)$.} 
\label{fig:smearing}
\end{figure}

\begin{figure}
\includegraphics[width=0.4\textwidth]{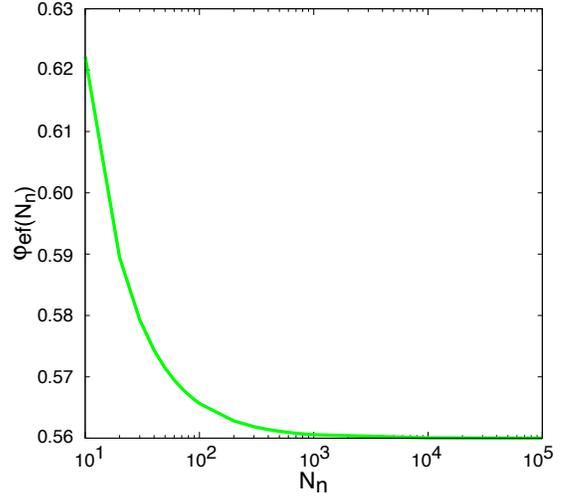}
\caption{ Convergence of the value of $\varphi _{ef}$ as the nodal numbers, $N _n$, increases. The parameters used were $\varphi  ^0_1 = 0.5$, $\varphi ^ 0 _2 = 0.62$ and $r=100$. From $N_n = 500$ on-wards we have that $\varphi _{ef} \approx (\varphi _1 + \varphi _2)/2=0.56$.} 
\label{fig:convergence}
\end{figure}

\noindent
To avoid the point at $x ^* = 1/2$ where the derivative of $\phi(x^*, t ^*=0)$ is undefined, the function $\varphi(x^*, t=0)$ was smoothed out according to:

\begin{equation}
\varphi _s (x^*) = \frac{\varphi _1 ^0 e^{r/2} + \varphi _2 ^0 e^{rx^*}}{e^{r/2} + e^{rx^*}},
\label{sigmoid_function}
\end{equation}

\noindent
where $r$ controls the slope of the function. In Fig. \ref{fig:smearing} the function $\varphi _s (x^*)$ is plotted according to Eq. (\ref{sigmoid_function}) for several values of $r$ with fixed $\varphi _1 ^0 = 0.5$, $\varphi _2 ^0 = 0.62$ as an example.  Fig. \ref{fig:convergence} shows, for the same $\varphi ^0 _1$ and $\varphi _2 ^0$ values but choosing $r = 100$, the convergence of the value of $\varphi _{ef} = \int _0 ^{1} \varphi _s(x^*)dx^* $ as the number of nodal points used in the discretization of Eq. \ref{Difussion_equation_eq} , $N _n$, increases. Notice that $\varphi _{ef} \approx (\varphi _1 ^0 + \varphi _2 ^0)/2$. From Fig. \ref{fig:convergence} we can conclude that a suitable choice of $N _n$, for the above parameters, is $N _n = 500$.

\noindent
To see how the system relaxes for $t >0$ for the given initial condition, we monitored the decay of the difference of the average of the densities in each side of the box defined by

\begin{equation}
\varphi ^* (t) = \frac{\braket{\varphi _2 (x^*,t^*)} - \braket{\varphi _1 (x^*,t^*)}}{\braket{\varphi _2 (x^*,t^*=0)} - \braket{\varphi _1 (x^*,t^*=0)}},
\end{equation}

\noindent
where

\begin{equation}
\braket{\varphi _1 (x^*,t^*)} =2 \int _0 ^{\frac{1}{2}} \varphi(x^*, t^*) dx^*
\end{equation}

\begin{equation}
\braket{\varphi _2 (x^*,t^*)} = 2 \int _{\frac{1}{2}} ^{1} \varphi(x^*, t^*) dx^*.
\end{equation}

\noindent so that a value of $\varphi ^* (t)=0$ indicates that a uniform value of the density has been achieved among the cells.\par

\begin{figure}
\includegraphics[width=0.4\textwidth]{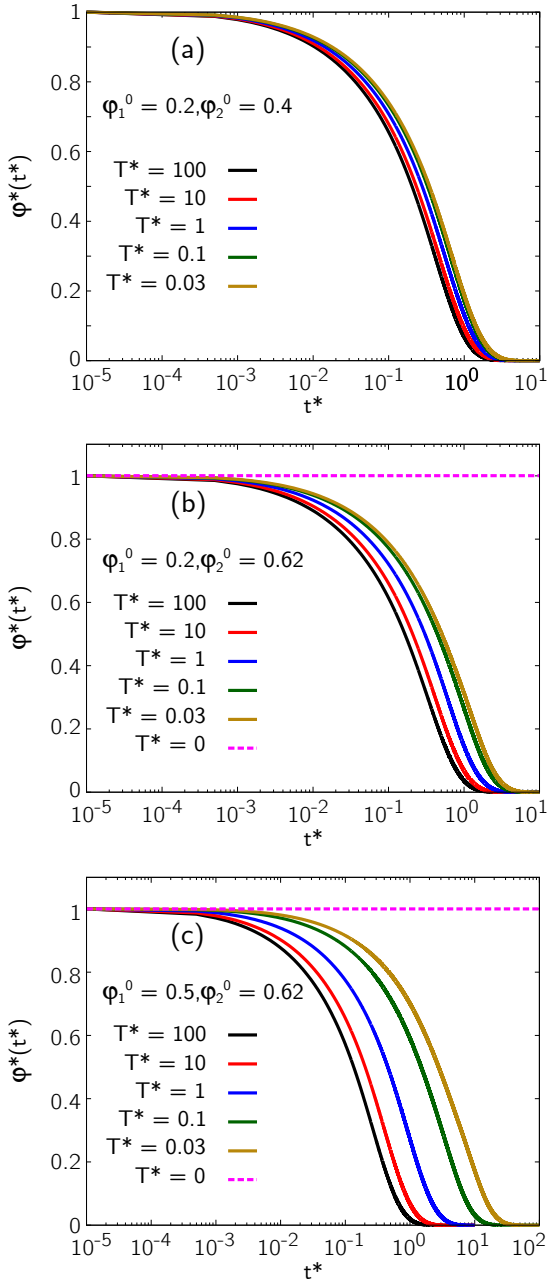}
\caption{ Relaxation of $\varphi ^* (t ^*)$ for different initial condition as function of the temperature $T^*$. (a) $\varphi ^0_1 = 0.2$, $\varphi  ^0_2 = 0.4$, (b) $\varphi ^0 _1 = 0.2$, $\varphi ^0 _2 = 0.62$ and (c) $\varphi ^0 _1 = 0.5$, $\varphi ^0 _2 = 0.62$.} 
\label{fig:Isotermas}
\end{figure}

Fig.\ref{fig:Isotermas} presents the time evolution of $\varphi ^*(t^*)$ for different pairs of the initial fractions $\varphi ^0 _1$ and $\varphi ^0 _2$ as well as for several values of $T ^*$. The mobility $b(\varphi (x ^*,t ^*))$, used in our calculations, was linearly interpolated between the equilibrium mobilities $b(\varphi_{i-1})$ and $b(\varphi _i)$, previously calculated with the NE-SCGLE framework, Eqs. (\ref{TimeEvolutionFk_eq})-(\ref{TimeEvolutionmovility_eq}),
for some $i = 1, \cdots, M$, being $M$ some integer. To find $i$, the condition $\varphi _{i-1} < \varphi (x ^*, t ^*) < \varphi _{i}$ must be satisfied; thus

\begin{equation}
b(\varphi(x^*,t ^*)) = b(\varphi _{i-1}) + \frac{b (\varphi _{i-1}) - b (\varphi _i)}{\varphi _{i-1} - \varphi _i} (\varphi(x^*,t ^*) - \varphi _{i-1}).
\end{equation}

\begin{figure}
\includegraphics[width=0.45\textwidth]{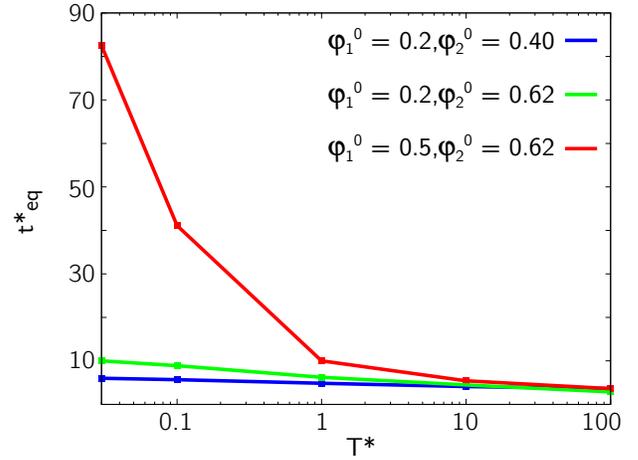}
\caption{Equilibration time $t ^* _{eq}$ as function of temperature $T ^*$ for the initial condition of Fig. \ref{fig:Isotermas}.} 
\label{fig:t_eq}
\end{figure}

\noindent
In general, we can see in Fig. \ref{fig:Isotermas} that $\varphi ^*(t^*)$ drop to zero (the cells become uniform) after a certain time for all the chosen parameters except when $T ^*=0$. Please notice that since the values of $T ^*$ are defined at the beginning of the simulation, physically this means that whatever the values of $\varphi ^0 _1$ and $\varphi ^0 _2$ are, the cells are driven instantaneously to the selected temperature at $t ^*=0$. When time starts to march, the zone with higher density and low mobility ($\varphi ^0 _2$) starts to diffuse towards the zone with lower densities and higher mobility until a uniform density is reached.
During this relaxation, the decrease in the density at $1/2<x ^* < 1$ induces an increase in the average mobility, whereas in the region $0<x ^* < 1/2$ occurs the inverse process, \textit{i.e.}, the mobility decreases as $\varphi _1 (x ^*, t^*)$ goes up. Also notice that an increase of the initial density gradient does not necessarily translates into a faster density relaxation, as one will expect due to an increase of the driving force (compare the curves between Figs.\ref{fig:Isotermas}(a) and (b) where, for the first case the initial volume fraction difference is 0.2 while for the latter is 0.42. This effect is due to the fact that the mobility of the particles is slower in the denser zone for the case denoted in Fig. \ref{fig:Isotermas}(b). The effect of the temperature, on the other hand, is to accelerate or delay the relaxation of $\varphi ^*(t ^*)$, as one can see in Fig. \ref{fig:Isotermas}: reduction of $T ^*$ yields a shift in the equilibration time, $t ^*_{eq}$, to a higher values, where $t ^* _{eq}$ is defined as the time that takes $\varphi ^* (t ^*)$ to reach zero, see also Fig. \ref{fig:t_eq}. This figure illustrates that for the diluted system composed by the pair $\varphi ^0_1 = 0.2$, $\varphi ^0_2 = 0.4$ (blue line), the shift in $t ^*_{eq}$ as a function of temperature is almost negligible. A similar trend happens in the pair $\varphi ^0 _1 = 0.2$, $\varphi ^0 _2 = 0.62$ in which the difference in $t ^* _{eq}$ is more evident but still small. In contrast, for the dense system having $\varphi ^0 _1 = 0.5$, $\varphi ^0 _2 = 0.62$ (red line), the magnitude of $t ^* _{eq}$ changes significantly as the reduced temperature increases up to a point ($T ^*\approx100$) where the curve converges with the diluted cases. This difference in diffusion between diluted and dense systems is due to the fact that the former, in overall, have greater mobility than the latter.\\

\noindent
Another characteristic that it is worth noticing in our systems is that the curves shown in Fig. \ref{fig:Isotermas} have the same shape and are only shifted horizontally for any value of $T ^*$ except zero as well as for any value of the initial density gradient. This highlights again the relevance of the mobility $b(\varphi (x^*,t^*))$ in the relaxation of the density $\varphi(x^*,t^*)$. In previous works using the NE-SCGLE theory, the mobility has shown to play an important role in finding stationary arrested states out of equilibrium  \cite{pedro2, LuisEnrique2013, olais, lira2019first, lira2021fundamental}; the value $b(\varphi (x^*,t^*)) = 0$ for example means that the particles localized at $x^*$ and being at time $t^*$ cannot longer diffuse, leading to a system trapped in an heterogeneous configuration even if non-zero density gradients are present. Hence, the system is found in a non-equilibrium arrested heterogeneous state. Given that we are considering an isothermal process attached to equilibrium (because we are using the equilibrium version of the NE-SCGLE theory to calculate $b (\varphi)$), for temperatures $T ^*$ lower than $T ^* _c (\varphi ^0 _2)$ the system will remain static as time marches with no particle diffusion. In Fig. \ref{fig:Isotermas} (b) and (c) the horizontal dotted pink lines represent the evolution of $\varphi ^*(t ^*)$ for $T ^* =0$. With Eq. (\ref{gamma}) we have determined that $T ^* _c (\varphi ^0_2 = 0.62) = 0.02$, thus, for this particular temperature, $T ^* < T^ * _c(\varphi ^0 _2)$ and the density of the configuration remains unchanged. Thus, in general, for a high heterogeneous system at temperature $T ^*$, no matter the size of the clusters  characterizing the initial heterogeneity, the clusters of density $\varphi(\mathbf{r},t)$ will relax only if $T^* _c (\varphi(\mathbf{r},t))<T^*$.

\section{Conclusions}\label{sec:conclusions}
In this work we have shown that the NE-SCGLE theory, a theory that studies structural relaxation, can be used to solve the spatial and temporal evolution of density heterogeneities including the possibility of dynamic arrest. In this way, the present theory can manage the full range of volume fractions, $\varphi(\mathbf{r},t)$, and reduced temperatures, $T^*$, conforming the dynamic phase diagram: from the low density, free-diffusing regime controlled by the short-time self-diffusion coefficient and in which all heterogeneities fade away (uniformity), up to the high density, low mobility regime controlled by the collective diffusion and in which dynamic arrest can take place. 

In order to proceed with the computations we used a model of interacting spherical soft colloids and a solution strategy that consists in computing the so-called non-uniform mobility, $b(\mathbf{r},t)$, and the mean volume fraction in two separate steps. In the first one we computed the mobility using the NE-SCGLE apparatus and which, in turn, uses the microestructural information of the colloidal model via correlation functions. The key assumption proposed here was that the noise or fluctuations around the transient values of the density are essentially the same as the values around the corresponding stationary value of $\bar{n} (\mathbf{r})$. Therefore, with the NE-SCGLE equations we can tabulate the values of $b(\mathbf{r},t)$ and use them as input in the diffusion equation via interpolation formulas.

In the second step we only require to focus efforts in solving the diffusion equation and handle the expression of the chemical potential which, again, recapitulates the microscopic information via the entropic and enthalpic part (particle interactions) of the total potential. After Taylor expanding the chemical state function around a uniform density value and assuming that the spatial gradients of the density fluctuations are small, we end up with an expression in terms of its value at the uniform density but corrected by the equilibrium structure factor which takes care of the cumulative many-body interaction of the particles. These terms are further collected in a generalized diffusion coefficient. The final form of the diffusion equation was solved using the finite element formulation (weak form) which has the benefit in reducing the degree of differentiation of the original equation and which can be further used as a basis to handle problems in different applications.

In order to show the efficacy of our method to solve the density relaxation, we studied in particular the density-decay of a initially non-uniform system (soft-spheres interacting with the truncated Lennard-Jones potential) under isothermal conditions. To illustrate how the density relaxes, we divided the system into two boxes each one having different initial densities and then monitor the time evolution of the difference between the average density of the two boxes. For a given set of initial density profiles, we observed that the system relaxes towards a uniform density in a time period named the equilibration time. The overall method showed consistency in the sense that this equilibration time increases as the temperature of the system decreases, as expected. We further observed that for the same value of $T^{*}$, a diluted system can diffuse faster than a denser system even if the latter has a higher initial density gradient: such an effect is a direct consequence of a lower mobility factor in dense systems. Finally, the present methodology can also capture instances ($T^{*}<T^{*}_{c}$) where the initial density gradient is accompanied by a very small mobility factor: in this case the initial heterogeneity is seen to be time-arrested.

Finally, it is worth saying that although our method to study the diffusion of glass forming colloids is apparently simple, there is still plenty of room to bring even closer these kind of theoretical developments to real industrial applications. The formalism presented here can be expanded easily to other more complicated situations than the isothermal case, for instance, considering cooling/heating rates acting over the system. Future efforts should be directed as well in extending the theory to consider non-uniform temperature landscapes (heat flow) as well as fluid flow. About the latter, we anticipate that flow not only changes the mean or stationary quantities such as the structure factor (case of slow or laminar flow) but also changes the way the density fluctuations relaxes (case of transient or turbulent flows).

\section*{Acknowledgements}
The authors acknowledge the financial support provided by CONACyT M\'exico through grants: C\'atedras CONACyT-1631 and CB-2015-01-257636. Besides we appreciate the technical assistance of Jose Limon and  the computational infrastructure provided by LANIMFE .

\appendix
\section{Numerical solution of  Eq. (\ref{Difussion_equation_eq})}\label{appendix}
In this appendix we will show a numerical scheme to solve Eq. (\ref{Difussion_equation_eq}) using the Finite Element formulation. First, lets write Eq. (\ref{Difussion_equation_eq}) in dimensionless form as follows:

\begin{equation}
\frac{\partial \varphi(\mathbf{r} ^*, t^*)}{\partial t^*} =  \nabla ^* \cdot [ b(\varphi(\mathbf{r}^*,t^*)) \varphi(\mathbf{r}^*, t^*) \nabla ^* \varphi(\mathbf{r}^*, t^*) ],
\label{Difussion_equation_local}
\end{equation}

\noindent where $t^* = t/\tau _M$, $\mathbf{r}^* = \mathbf{r}/L$, $L=V ^{1/3}$ is characteristic length and $\tau _M$ is the macroscopic characteristic time defined by $\tau _M = L ^2 /D ^* _0$. Physically, $\tau _M$ is the time required by a tagged particle to diffuse the available volume.

Eq. (\ref{Difussion_equation_local}) is a non-linear second order partial  differential equation. The most easy, fast and general way for solving Eq. (\ref{Difussion_equation_local}) is using numerical methods such as finite differences or finite element. In the finite element approach, the variational formulation (weak form) of Eq. (\ref{Difussion_equation_local}) in one dimension is given by\cite{larson2013finite}

\begin{multline}
\int _0 ^1 \frac{\partial \varphi (x^*,t^*)}{\partial t^*} \varphi _i (x^*) dx^* = \bigg[ b(\varphi(x^*,t^*))\varphi(x^*,t^*)\varphi _i(x^*)\bigg]_0 ^1 \\
- \int _0 ^1 b(\varphi(x^*,t^*))\varphi(x^*,t^*)\frac{\partial \varphi (x^*,t^*)}{\partial x^*}\frac{d \varphi _i (x^*)}{dx^*}dx^*
\label{variational_form}
\end{multline}

\noindent where the region of integration $\mathcal{I} = [0,1]$ is a domain where the function $\phi(\mathbf{x^*},t)$ is piecewise defined and  $\varphi _i (x)$ are the test functions in one dimension defined in the $i$-interval of a partition of $\mathcal{I}$\cite{larson2013finite}.  Although the test functions, $\varphi _i (x)$, are in general non orthogonal, we can write any continuous piece-wise linear function by constructing an expansion based on $\varphi _i (x)$. Thus,

\begin{equation}
\varphi (x^*,t^*) = \sum _{j=0} ^{N_n -1} \eta _j(t^*) \varphi _j(x^*),
\end{equation}

\noindent
where $N _n$ are the total number of nodes in the partition of $\mathcal{I}$. Substituting the expansion of $\varphi (x^*,t^*)$ and  considering that there is no flux of matter at the outer boundaries, \textit{i.e.}, the first term of the r.h.s. of Eq. (\ref{variational_form}) is zero, hence,

\begin{multline}
\sum _{j = 0} ^{N_n -1} \frac{d \eta_j (t^*)}{dt^*} \int _0 ^1 \varphi _j (x^*) \varphi _i (x^*)dx^* = \\ - \sum _{j = 0} ^{N_n -1} \eta _j (t^*) \int _0 ^1 b(n(x^*,t^*))\varphi (x^*,t^*)\frac{d \varphi_j (x^*)}{dx^*}\frac{d \varphi _i(x^*)}{dx^*}dx^*,
\end{multline}

\noindent where we have only used the expansion of $\varphi (x^*,t^*)$ in the derivative of the r. h. s. of Eq. (\ref{variational_form}) since we have a non-linear differential equation\cite{larson2013finite}. In matrix form the last equation can be written as

\begin{equation}
A \dot{\eta}(t) = -B(t)\eta (t)
\label{matrix_form}
\end{equation}

where

\begin{equation}
A _{ij} = \int _0 ^{1} \varphi _i(x^*) \varphi _j(x^*) dx^*,
\label{A_1d}
\end{equation}

and

\begin{equation}
B _{ij}(t) = \int _0 ^1 b(\varphi(x^*,t^*))\phi(x^*,t^*)\frac{d \varphi_i (x^*)}{dx^*}\frac{d \varphi _j(x^*)}{dx^*}dx^*, 
\label{B_1d}
\end{equation}

\noindent
with  $i,j = 1, \cdots, N_n -1$. In order to obtain the discrete time form of Eq. (\ref{matrix_form}) we can make $\dot{\eta} (t)\approx (\eta_{l+1}  - \eta _{l})/ \Delta t$, where $\eta _l = \eta (t_0 + l \Delta t)$, with $l = 0,1...,$ being $\Delta t$ the time step and $t _0 = 0$. Applying this approximation for $\dot{\eta}(t)$ and doing $B _{l}=B _{l-1}$ (it is understood that $B _l  = B (t _0 + l\Delta t)$), the discrete form of Eq. (\ref{matrix_form}) ends in the following form:

\begin{equation}
\bigg[ A + \Delta t B_{l} \bigg] \eta _{l +1} = A \eta _{l}.
\label{discrete_form}
\end{equation}

\bibliography{HeterogeneitiesPaper}

\end{document}